%% file: programoffer1.tex
\documentclass[11pt]{article}
\usepackage{amsfonts,amsmath,amssymb,amsthm}
\usepackage[pdftex]{graphicx}
\usepackage[hmargin=1in,vmargin=1in]{geometry}
\usepackage{natbib}
\usepackage[colorlinks,citecolor=blue]{hyperref}
\usepackage{enumerate}
\usepackage{caption}
\usepackage[usenames,dvipsnames]{xcolor}
\usepackage{footmisc,fancyvrb} 
\usepackage{xfrac} 
\usepackage{cleveref}
\usepackage{multirow}
\usepackage{threeparttable}
\usepackage{array}
\newcolumntype{C}[1]{>{\centering\let\newline\\\arraybackslash\hspace{0pt}}m{#1}}
\newcolumntype{L}[1]{>{\flushleft\let\newline\\\arraybackslash\hspace{0pt}}m{#1}}
\newcolumntype{R}[1]{>{\flushright\let\newline\\\arraybackslash\hspace{0pt}}m{#1}}

\usepackage{xr}
\VerbatimFootnotes
\def\qed{\rule{2mm}{2mm}}

\newcommand\ddfrac[2]{\frac{\displaystyle #1}{\displaystyle #2}}
\DeclareMathOperator*{\argmax}{arg\,max}

\usepackage{tikz,graphicx,pgfplots,pgfkeys,subcaption}   

\def\addlegendimage{\csname pgfplots@addlegendimage\endcsname}
\usepackage{booktabs}
\usepackage[pagewise,mathlines]{lineno}
\mathchardef\dash="2D

\theoremstyle{definition}

\newtheorem{proposition}{Proposition}

\newtheorem{assump}{Assumption}
\newenvironment{assumption2}[2][]
  {\renewcommand\theassump{#2}\begin{assump}[#1]}
  {\end{assump}}
\newtheorem{example}{Example}

\usepackage{etoolbox} 
\AtEndEnvironment{remark}{~\qed}%
\AtEndEnvironment{example}{~\qed}%

\begin{document}

\author{
Vishal Kamat \\
Toulouse School of Economics\\
University of Toulouse Capitole\\
\url{vishal.kamat@tse-fr.eu} \vspace{-1ex}
}

\title{ \vspace{-7ex}  Identifying the Effects of a Program Offer with an Application to Head Start\thanks{I am extremely grateful to Ivan Canay, Chuck Manski and Alex Torgovitsky for their extensive guidance and feedback. I also thank the editor, two anonymous referees, Eric Auerbach, Gideon Bornstein, Pedro Carneiro, Joel Horowitz, Tom Meling, Magne Mogstad, Francesca Molinari, Sam Norris, Matt Notowidigdo, Nicolas Inostroza, Rob Porter, Pedro Sant'Anna, Lola Segura, Azeem Shaikh, Max Tabord-Meehan and conference and seminar participants at several institutions for useful comments, and Research Connections for providing data on the Head Start Impact Study. Funding from ANR under grant ANR-17-EURE-0010 (Investissements d'Avenir program) and the Robert Eisner Memorial Fellowship is gratefully acknowledged. First version: arXiv:1711.02048 dated November 6, 2017.}} 

\maketitle

\vspace{-1.05cm}
\begin{abstract}
I propose a treatment selection model that introduces unobserved heterogeneity in both choice sets and preferences to evaluate the average effects of a program offer. I show how to exploit the model structure to define parameters capturing these effects and then computationally characterize their identified sets under instrumental variable variation in choice sets. I illustrate these tools by analyzing the effects of providing an offer to the Head Start preschool program using data from the Head Start Impact Study. I find that such a policy affects a large number of children who take up the offer, and that they subsequently have positive effects on test scores. These effects arise from children who do not have any preschool as an outside option. A cost-benefit analysis reveals that the earning benefits associated with the test score gains can be large and outweigh the net costs associated with offer take up.
\end{abstract}

\thispagestyle{empty}



\newpage
\setcounter{page}{1}

\section{Introduction}\label{sec:introduction}

A central task of program evaluation is to evaluate policies relevant to a program. For many programs, the baseline policy corresponds to providing individuals with an offer to participate in the program. For example, in a preschool program, children may be provided with an offer to preschools in the program, while in a microfinance program, individuals may be provided with a loan offer. The evaluation task for such programs then translates to evaluating the effects of providing an offer. Indeed, evaluating the number of individuals who may take up the offer and the outcomes they later experience allows one to assess the costs and benefits of providing an offer and, in turn, the returns to the policy.

To analyze such policy relevant questions, the standard approach in the literature is to model how individuals select into the program and exploit the structure of the model to define parameters capturing the policy question of interest \citep{heckman/vytlacil:07}. As highlighted in Section \ref{sec:comparison}, the typical model to do so corresponds to individuals choosing by maximizing utility over the various alternatives, where a single dimension of unobserved heterogeneity governs the utility under each alternative. However, as a program offer corresponds to providing the program in an individual's choice set, analyzing the effects of an offer naturally requires a richer model that distinguishes the role of choice sets and preferences in determining utility.

In this paper, I propose a novel treatment selection model that introduces unobserved heterogeneity in both choice sets and preferences to learn about the effects of providing a program offer. By distinguishing the role that choice sets and preferences play, the richer structure provides a building block to define a range of parameters evaluating the effects of a program offer that correspond to comparing participation decisions and outcomes under counterfactual choice sets with and without the program.

Having defined these parameters, I then show how to exploit the model structure to learn about them in the presence of instrumental variables that induce exogenous variation in choice sets. I allow the model to be nonparametric and the dependence between choices sets, preferences and outcomes to be unrestricted. These features imply that the parameters are generally partially identified. Deriving sharp analytical bounds for these parameters, however, can be difficult given the rich model structure and geometry of the parameters. In turn, by leveraging results from \cite{charnes/cooper:62}, I illustrate how a linear programming procedure can be used to computationally characterize the identified sets. The procedure flexibly applies to the various parameters and allows a range of shape restrictions to be imposed on the baseline model.

As highlighted in Section \ref{sec:examples}, the developed tools can be applied to a number of setups, specifically randomized experiments, where exogenous variation in choice sets naturally arises. For example, in the Head Start Impact Study (HSIS), families were randomly assigned to a treatment and control group, where those in the treatment group were more likely to be assigned a Head start preschool offer than those in the control group. Similarly, in a microfinance experiment in \cite{angelucci/etal:15}, individuals in treatment villages were more likely to receive loan offer than those in control villages; while in the Oregon Health Field Experiment, those in the treatment group where more likely to receive Medicaid than those in the control group.

To empirically illustrate the tools, I apply it to evaluate the effects of providing a Head Start offer using data from the HSIS. Previous analyses have focused on the instrumental variable estimand that evaluates effects for the subgroup of compliers, and their variants that account for where these children would have enrolled in the control group \citep{feller/etal:16, kline/walters:16, puma_etal:10}. As highlighted in Section \ref{sec:setting}, the complier subpopulation corresponds to the policy relevant one when the policy of interest corresponds to the variation induced by the experiment. In turn, they leave open the question on the effects for the subgroup affected in a more general policy providing offers to Head Start.

The application reveals several conclusions. First, providing an offer to Head Start affects a large percentage of children who take up the offer and participate in Head Start, and can significantly improve their test scores. In particular, 82.5$\%$ of children who are provided an offer choose to take it up, and their subsequent tests scores can improve on average between 0.21 and 0.50 standard deviation points. Second, these effects primarily arise from the subgroup of children who do not have an offer to an alternative (non-Head Start) preschool and hence any outside option. Finally, using the estimates to perform a cost-benefit analysis of the policy, the earning benefits associated with the test score gains can potentially be large and outweigh the net costs associated with the take up of the offer. Specifically, when the relation between test scores and earnings is towards the higher end of estimates noted in the literature, the average benefits net of costs per child can range between $\$$12,365 and $\$$37,873.

The developed tools contribute to a large literature that exploit instrumental variables to learn about treatment effects and, specifically, in setups that permit multiple treatments. One strand of the literature focuses on the instrumental variable estimand and shows how it can evaluate treatment effects local to various complier subgroups under different types of monotonicity conditions \citep[e.g.,][]{angrist/imbens:95,heckman/pinto:18,kline/walters:16, kirkeboen/etal:16,pinto:19}. Indeed, in cases where the instrument itself corresponds to the policy of interest, the compliers correspond to those affected by the policy change. But, for other more general policies such as the provision of a program offer, one may require learning about treatment effects for various other subgroups of interest. 

To this end, the analysis in this paper is more closely related to an alternative strand of the literature that considers choice models to define and learn about various policy relevant parameters that move beyond local effects \citep[e.g.,][]{heckman/etal:06,heckman/etal:08,heckman/vytlacil:07,kline/walters:16,lee/salanie:19}. The proposed choice model complements those in these papers that are each tailored for different questions of interest. Specifically, unlike these papers, the model here introduces structure to distinguish the role of choice sets and preferences in selection, which plays an essential role in defining and analyzing the effects of a program offer.

In developing the selection model, the analysis exploits elements from the literature on discrete choice analysis. As highlighted in Section \ref{sec:comparison}, \cite{manski:07} proposed a choice model with heterogeneity in both choice sets and preferences where one only assumed individuals to have a strict preference relationship over the various alternatives. However, while this model allowed for unobserved preferences, it took the choice sets to be observed and exogenously vary. Designed for setups more commonly arising in the program evaluation context, the model in this paper builds on this model to allow for unobserved choice sets and instrumental variable variation in choice sets. In this sense, the model also contributes to a growing literature on  discrete choice analysis under unobserved choice sets \citep[e.g.,][]{abaluck/adams:18,barseghyan/etal:18,cattaneo/etal:18,crawford/etal:19}, where models with alternative structure and assumptions between choice sets and preferences have been proposed.

The remainder of the paper is organized as follows. Section \ref{sec:setup} introduces the model, defines the average effects of an offer, and illustrates how various empirical examples fall in this setup. Section \ref{sec:computation} shows how to learn about the parameters of interest. Section \ref{sec:empirical_results} applies the tools to the HSIS data. Section \ref{sec:conclusion} concludes. Proofs and auxiliary results complementing the main analysis are presented in the Supplementary Appendix.

\section{Setup}\label{sec:setup}

\subsection{Model}\label{sec:model}

For each individual, we observe an outcome of interest $Y$, their received treatment $D$, and their assigned instrument value $Z$. We take the list of of possible treatments to be given by a discrete set $\mathcal{D}$, and also model the support of the instrument to be given by a discrete set $\mathcal{Z}$. As usual, I assume the observed outcome to be generated by the potential outcomes structure. Denoting by $Y(d)$ the potential outcome had the individual received treatment $d \in \mathcal{D}$, the observed outcome is given by
\begin{align}\label{eq:Y_equation}
  Y = \sum_{d \in \mathcal{D}} Y(d) 1\{D=d\} \equiv Y(D)~.
\end{align}

In addition, I assume the received treatment to be generated by a selection model that introduces heterogeneity in both choice sets and preferences. Specifically, let $C \in \mathcal{C}$ denote the individual's received choice set that corresponds to a non-empty subset of $\mathcal{D}$ to where the individual receives offers from and has the option to participate, and let $U(d) \in \mathbf{R}$ denote the individual's utility from participating in alternative $d \in \mathcal{D}$. The observed treatment is then given by the treatment that maximizes utility across the alternatives present in their realized choice set as follows
\begin{align}\label{eq:utility_max}
  D = \argmax_{d \in \mathcal{C}} U(d)~.
\end{align}

Since the utilities do not possess any cardinal value, different monotonic transformations of them will generate observationally equivalent choices. It is therefore useful to directly refer to the underlying preference type that they represent. To this end, assuming individuals have a strict preference relation over the set of alternatives, let $U$ denote the individual's preference type corresponding to a strict preference relationship over $\mathcal{D}$, and let $\mathcal{U}$ denote the set of such types.\footnote{For example, if $\mathcal{D} = \{1,2\}$, the set of preference types can be given by $\mathcal{U} = \{ (1 \succ 2),~(2 \succ 1)\}.$} In addition, let $d(u,c)$ denote the known choice function that corresponds to what preference type $u \in \mathcal{U}$ would choose under a non-empty subset $c \subseteq \mathcal{D}$, i.e. under a given feasible choice set.\footnote{For example, $d((2 \succ 1), \{2,1\}) = 2$ as preference type $(2 \succ 1)$ prefers $2$ to $1$ and hence would choose $2$ when faced with the choice set containing $2$ and $1$.} Using this notation, the utility maximization relationship in \eqref{eq:utility_max} can be alternatively re-written in terms of the preference type and realized choice set by
\begin{align}\label{eq:selection_equation}
 D = \sum_{u \in \mathcal{U} , c \in \mathcal{C}} d(u,c) I\{U = u, C = c\} \equiv d(U,C)~.
\end{align}

I also assume the instrument to affect where an individual receives an offer from and hence their choice set, while being excluded from how it affects utilities and outcomes. To this end, using the potential outcomes structure again, let $C(z)$ denote the individual's potential choice set had they been assigned instrument value $z \in \mathcal{Z}$, and let the realized choice set be given by
\begin{align}\label{eq:C_equation}
  C = C(Z)~.
\end{align}
As illustrated in Section \ref{sec:examples}, the relation between $C(z)$ across $z \in \mathcal{Z}$ will allow imposing how the instrument varies the provision of an offer.

Finally, as usual, I take the instrument to be exogenous and statistically independent of the underlying unobserved variables as follows.

\begin{assumption2}{E}\label{ass:E}
(Exogeneity) $(\{Y(d) : d \in \mathcal{D} \},U,\{C(z) :  \in \mathcal{Z}\}) \perp Z~.$
\end{assumption2}

\subsection{Comparison to a Standard Choice Model}\label{sec:comparison}

A standard choice model in the literature on treatment effect analysis is to take the observed choice to be given by
\begin{align*}
  D = \sum_{z \in \mathcal{Z}} D(z) 1\{Z=z\}~,
\end{align*}
where
\begin{align}\label{eq:standard}
 D(z) = \argmax_{d \in \mathcal{D}} \bar{U}(d,z)~
\end{align}
for $z \in \mathcal{Z}$ with $\bar{U}(d,z)$ corresponding to the utility under $d \in \mathcal{D}$ \citep{heckman/vytlacil:07}. This choice model is indeed more agnostic about the choice process than that in \eqref{eq:utility_max} and can nest it by taking $\bar{U}(d,z) = U(d)$ if $d \in C(z)$ and $\bar{U}(d,z) = - \infty$ if $d \notin C(z)$ for each $d \in \mathcal{D}$ and $z \in \mathcal{Z}$, i.e. by taking utility to be equal to that of enrolling in that alternative if it is in the choice set and to be equal to negative infinity if not as that alternative cannot be feasibly chosen. In other words, the utility here can be interpreted as combining that of choosing an alternative and whether that alternative is in the choice set into a single dimension.

However, by not separating the heterogeneity in preferences and choice sets, this model does not straightforwardly allow analyzing the effects of providing an offer in comparison to the richer structure of the model in \eqref{eq:utility_max}. Importantly, this richness allows modeling the concept of an individual receiving an offer to an alternative or not---namely, by taking that alternative to be present in their choice set or not. As we will observe below, this plays an essential role in defining and then learning about a range of effects of providing a program offer by comparing responses under counterfactual choice sets with and without the program.

Explicitly modeling the role of choice sets is a common practice in the literature on discrete choice analysis when one wants to analyze choice under counterfactual choice sets. To this end, it is useful to highlight that the proposed model draws on this conceptual connection and, in particular, to the model in \cite{manski:07} with some novel modifications---see also the related analysis of \cite{marschak:59}. \cite{manski:07} similarly introduced for each individual an unobserved preference type and a choice set which together determine observed choice as in \eqref{eq:selection_equation}. However, designed with alternative empirical setups in mind, the model assumed the choice set $C$ to be observed and statistically independent of preferences, i.e. $C \perp U$. In the program evaluation setup studied in this paper, this amounts to assuming that from where offers were received are observed and independent of the individual's preferences, both of which may generally not be valid. The proposed model in turn shows how to alternatively consider a choice equation with unobserved choice set and instrumental variable variation in choice sets.

\subsection{Examples}\label{sec:examples}

Before proceeding to show how we can define the effects of an offer in the proposed model, I illustrate how various empirical examples fit in the model. I focus on various randomized experiments that naturally induce exogenous variation in choice sets. In each experiment, I highlight what the treatment $D$ and instrument $Z$ correspond to and how the experiment varied provision of offers through the relation between $C(z)$ across $z \in \mathcal{Z}$.

\begin{example}{(Head Start Impact Study)}\label{ex:hsis}
In this experiment, children were choosing between various preschools and the experiment randomly assigned children to a treatment or control group, where those in the treatment group were guaranteed an offer to preschools in the Head Start program \citep{puma_etal:10}. In this case, $D$ corresponds to a child's choice of preschool, where we can take $\mathcal{D} = \{p,a,n\}$ such that $p$ denotes preschools part of the Head Start program, $a$ denotes alternative preschools not part of the program, and $n$ denotes the choice of no preschool; and $Z$ corresponds to an indicator for their assigned group, where $Z=0$ denotes that the individual is in the control group while $Z=1$ denotes that they are in treatment group. 

While the experiment ensured an offer to Head Start preschools to individuals in the treatment group, those in the control group were not prevented from receiving an offer from sources other than the experiment. In turn, we can assume
\begin{align}\label{eq:choiceset_hsis}
  C(1) = C(0) \cup \{p\}~,
\end{align}
i.e. treatment alters the choice set by solely guaranteeing that Head Start preschools are in it, where $C(0),C(1) \in \mathcal{C} = \{c \subseteq \mathcal{D} : n \in c\}$ as the outside option of no preschool is always in the choice set.
\end{example}

\begin{example}{(Microfinance experiment)}
In a microfinance experiment conducted in \cite{angelucci/etal:15}, individuals where choosing what type of loan product to choose and the experiment randomly assigned individual living certain areas to a treatment or control group, where those in treated areas were guaranteed a loan offer from the Credit Mujer program. In this case, $D$ corresponds to the individual's choice of loan, where we can take $\mathcal{D} = \{p,a,pa,n\}$ such that $p$ denotes a loan from the Credit Mujer program, $a$ denotes a loan not part of the program, $pa$ denotes loans from both the program and alternative sources and $n$ denotes the choice of no loan; and $Z$ corresponds to an indicator for their assigned group, where $Z=0$ denotes that the individual lives in a control area and $Z=1$ denotes that they live in a treatment area. 

Similar to the Head Start experiment above, while the experiment ensured an offer to Credit Mujer to individuals in treated areas, those in the control areas could potentially receive an offer from a treated area if, for example, they had a viable address there. In turn, we can again assume
\begin{align}\label{eq:choiceset_micro}
  C(1) = C(0) \cup \{p\}~,
\end{align}
i.e. treatment alters the choice set by solely guaranteeing that program is in it, where again $C(0),C(1) \in \mathcal{C} = \{c \subseteq \mathcal{D} : n \in c\}$. Here we also have the following additional logical restriction on choice sets
\begin{align}
  C(z) \cup \{pa\} = C(z) \cup \{p,a\}
\end{align}
for $z \in \mathcal{Z}$, which captures the logical fact that if Credit Mujer and alternative loan products are in the choice set then so should each one of them.
\end{example}

\begin{example}{(Oregon Health Field Experiment)}
In this insurance experiment, individuals were choosing between various insurance options and a lottery randomly assigned individuals to a treatment or control group, where those in the treatment group were guaranteed an offer to Medicaid conditional on satisfying eligibility requirements \citep{finkelstein/etal:12}. In this case, $D$ corresponds to an individual's insurance choice, where we can take $\mathcal{D} = \{p,a,n\}$ such that $p$ denotes Medicaid program insurance, $a$ denotes insurance from alternative non-Medicaid providers, and $n$ denotes the choice of no insurance; and $Z$ corresponds to an indicator for their assigned group, where $Z=0$ denotes that the individual is in the control group while $Z=1$ denotes that they are in treatment group. 

In this case, the lottery did not guarantee Medicaid to the treated individuals, but only made it more likely to potentially receive conditional on being eligible. In turn, in contrast to the above two examples, we can assume that
\begin{align}\label{eq:choiceset_ohfe}
  C(1) = C(0) \text{ or } C(1) = C(0) \cup \{p\} 
\end{align}
i.e. treatment either does not alter the choice set or does so by introducing the program in it, where as above $C(0),C(1) \in \mathcal{C} = \{c \subseteq \mathcal{D} : n \in c\}$.
\end{example}

\subsection{Defining Average Effects of an Offer}\label{sec:parameters}

I conclude this section by illustrating how to define various effects of an offer in the proposed model. To this end, recall that an offer to an alternative corresponds to an individual receiving that alternative in their choice set. The effects of an offer can therefore be defined by comparing potential responses under counterfactual choice sets with and without the alternative. To define these effects, it is useful to introduce additional notation to refer to the potential decisions and outcomes under a given choice set. Had the individual's choice set been pre-specified to a non-empty subset $c \subseteq \mathcal{D}$, let $D_{c} = d(U,c)$ denote the alternative they would choice from $c$ and $Y_{c} = Y(D_c)$ the subsequent outcome they would then earn.

For each individual, if an offer to an alternative of interest $d \in \mathcal{D}$, typically corresponding to the program, was provided then their resulting choice set would be given by $C_{+d} = C \cup \{d\}$, while if it was not then it would be given by $C_{-d} = C \setminus \{d\}$. The average (treatment) effect (ATE) of providing an offer to $d \in \mathcal{D}$ on choosing this alternative and subsequent outcomes can then be defined by taking the average value of the difference in the respective potential responses under these two choice sets given by
\begin{align}
 \text{ATE}^{D}(C_{+d},C_{-d}) &= E\left[1\{D_{C_{+d}} = d\} - 1\{D_{C_{-d}} = d\}\right]~, \label{eq:ATE_D_temp} \\
 \text{ATE}^{Y}(C_{+d},C_{-d}) &= E\left[Y_{C_{+d}} - Y_{C_{-d}}\right]~. \label{eq:ATE_Y}
\end{align}
Note that by construction we have $1\{D_{C_{-d}} = d\} = 0$, which captures the fact that if the individual is not provided an offer to an alternative then they naturally cannot choose it. As a result, observe that the former parameter can more simply be rewritten as
\begin{align}
 \text{ATE}^{D}(C_{+d},C_{-d}) = \text{Prob}\{D_{C_{+d}} = d\} \label{eq:ATE_D}
\end{align}
i.e. the proportion who choose the alternative when provided an offer. Observe that providing an offer does not affect the subgroup who do not take up the offer, i.e. those with $D_{C_{+d}} \neq d$, as for these children we have $Y_{C_{+d}} = Y_{C_{-d}}$. It is therefore useful to also define the average (treatment) effect of providing an offer conditional on participation (ATOP), i.e. for those who take it up and are hence affected, by 
\begin{align}
  \text{ATOP}(C_{+d},C_{-d}) = E\left[Y_{C_{+d}} - Y_{C_{-d}} | D_{C_{+d}} = d\right]~. \label{eq:ATOP}
\end{align}

We can also consider the above average effects of an offer conditional on a subgroup of interest. For example, as done in the empirical analysis in Section \ref{sec:alternative}, we may be interested in evaluating the heterogeneity in the effects of an offer to an alternative based on whether individuals do or do not have an offer to another alternative. In particular, we can define
\begin{align}
 \text{ATE}^{D}_{+d'}(C_{+d},C_{-d}) &= \text{Prob}\{D_{C_{+d}} = d | d' \in C\}~, \label{eq:ATE_D_a+} \\
 \text{ATOP}_{+d'}(C_{+d},C_{-d}) &= E[Y_{C_{+d}} - Y_{C_{-d}} |  D_{C_{+d}} = d, d' \in C] \label{eq:ATOP_a+}
\end{align}
to capture the proportion who take up the offer and the effects on their subsequent outcomes conditional on having an offer to alternative $d' \in \mathcal{D}$, or 
\begin{align}
 \text{ATE}^{D}_{-d'}(C_{+d},C_{-d}) &= \text{Prob}\{D_{C_{+d}} = d | d' \notin C\}~, \label{eq:ATE_D_a-} \\
 \text{ATOP}_{-d}(C_{+d},C_{-d}) &= E[Y_{C_{+d}} - Y_{C_{-d}} |  D_{C_{+d}} = d, d' \notin C] \label{eq:ATOP_a-}
\end{align}
to capture the analogous effects conditional on not having an offer to alternative $d' \in \mathcal{D}$.

\section{Computing the Identified Set}\label{sec:computation}

In this section, I show how to exploit the structure of the model and variation in choice sets to learn about the parameters of interest. For these purposes, it is useful to first introduce the underlying primitive of the model on which the analysis is based. Let $W = (\{Y(d) : d \in \mathcal{D} \},U,\{C(z) :  \in \mathcal{Z}\})$ summarize the latent variables for each individual. The primitive $Q$ then corresponds to the distribution of $W$. To ensure that $Q$ is a finite-dimensional object, I assume that the outcomes take values in a known discrete set $\mathcal{Y} \equiv \{ y_1, \ldots, y_M  \}$. This implies that $Q$ is a probability mass function with support contained in a discrete space $\mathcal{W} \equiv \mathcal{Y}^{|\mathcal{D}|} \times \mathcal{U} \times \mathcal{C}^{|\mathcal{Z}|}$, i.e. $Q : \mathcal{W} \to [0,1]$ and $\sum\limits_{w \in \mathcal{W}} Q(w) = 1$. As we will observe below, this allows characterizing what we can learn about the parameter using finite-dimensional computational programs.

\subsection{Identification Problem}\label{sec:identifiedset}

The objective is to learn about a pre-specified parameter of interest defined in Section \ref{sec:parameters}. Each of these parameters can be written in terms of a known function of $Q$. Denoting by $\theta(Q)$ a generic function for the pre-specified parameter, $\theta$ generally has the following form 
\begin{align}\label{eq:Qtheta}
   \theta(Q) = \ddfrac{ \sum_{w \in \mathcal{W}} a_{\text{num}}(w) \cdot Q(w) }{ \sum_{w \in \mathcal{W}} a_{\text{den}}(w) \cdot Q(w) }~,
\end{align}
where $a_{\text{num}} : \mathcal{W} \to \mathbf{R}$ and $a_{\text{den}} : \mathcal{W} \to \mathbf{R}$ are known functions, i.e. as a fraction of linear functions of $Q$. For example, for the parameter \eqref{eq:ATE_D} in the context of Example \ref{ex:hsis} when $d=p$, we can write it as a linear function, a special case of linear-fractional functions where the denominator takes a value of one by taking $a_{\text{den}}(w) = 1$ for all $w \in \mathcal{W}$, as follows
\begin{align}\label{eq:ATE_D_Q}
  \text{ATE}^D(C_{+p},C_{-p})(Q) =  \sum_{w \in \mathcal{W}^p}  \cdot Q(w)~,
\end{align}
where $\mathcal{W}^p = \{ w \in \mathcal{W} : d(u,c(1))  = p\}$. As illustrated in Appendix \ref{S-sec:rewriting_theta}, we can similarly do so for the remaining parameters of interest. Indeed, note that the parameters that do not condition on any event such as that in \eqref{eq:ATE_D} can be written as linear parameters, while those that condition on some event can be written more generally as linear-fractional parameters.

Given that the function $\theta$ is known, what we can learn about the parameter depends on the possible values that $Q$ can take. From the observed data and its relation to the latent variables through the structure \eqref{eq:Y_equation}-\eqref{eq:C_equation} along with Assumption \ref{ass:E}, we have that
\begin{align}\label{eq:Qdata}
 \sum_{w \in \mathcal{W}_{x}} Q_z(w) = \text{Prob}\{Y=y,D=d|Z=z\}
\end{align}
for all $x=(y,d,z) \in \mathcal{Y} \times \mathcal{D} \times \mathcal{Z} \equiv \mathcal{X}$, where $\mathcal{W}_{x}$ is the set of all $w \equiv (\{y(d'):d' \in \mathcal{D}\},u,\{c(z') : z' \in \mathcal{Z}\})$ in $\mathcal{W}$ such that $d(u,c(z))=d$ and $y = y(d)$. For the purposes of identification, the data moments on the right hand side of the above equation are assumed to be known with certainty. I discuss estimation and inference in Section \ref{sec:inference} below. In addition to the data, the assumptions we impose as in Section \ref{sec:examples} on the variation in choice sets also restrict $Q$. In particular, these restrictions can generally be stated as
\begin{align}\label{eq:Qass}
  \sum_{w \in \mathcal{W}_s} Q(w) = 0~,
\end{align}
where $\mathcal{W}_s$ is some known subset of $\mathcal{W}$, i.e. they impose zero probability on the occurrence of some events. For example, observe that \eqref{eq:choiceset_hsis} can be written as \eqref{eq:Qass} by taking $\mathcal{W}_s = \{w \in \mathcal{W} : c(1) \neq c(0) \cup \{p\}\}$. The restrictions in \eqref{eq:choiceset_micro}-\eqref{eq:choiceset_ohfe} can be similarly stated in this manner. Moreover, as presented in Section \ref{sec:estimates} in context of the empirical application, \eqref{eq:Qass} can also encompass additional such restrictions on the relation between outcomes and preferences.

Denoting by $\mathbf{Q}_{\mathcal{W}}$ the set of all probability mass functions on the sample space $\mathcal{W}$ and by $\mathbf{Q} = \{Q \in \mathbf{Q}_{\mathcal{W}} : Q \text{ satisfies } \eqref{eq:Qdata} \text{ and } \eqref{eq:Qass}\}$ the subset of $\mathbf{Q}_{\mathcal{W}}$ that satisfies the various restrictions imposed by the data and assumptions, what we can learn about the parameter can then be defined by the identified set,
\begin{align}\label{eq:identified_set}
 \Theta = \{ \theta_0 \in \mathbf{R} : \theta(Q) = \theta_0 \text{ for some } Q \in \mathbf{Q}  \}~,
\end{align}
i.e. the image of the space of admissible primitives under $\theta$. The objective of the analysis then translates to show how to compute the identified set.

\subsection{Characterization}\label{sec:characterization}

A standard approach in the literature is to analytically characterize the identified set---see \cite{manski:09} for a textbook treatment. While analytic bounds can provide intuition into what is driving identification, doing so in the above setup can be difficult due to the complicated model structure and parameter function $\theta(Q)$. To this end, in the following proposition, I illustrate how a linear programming procedure can be used to instead computationally characterize it---see, for example, \cite{balke/pearl:97}, \cite{demuynck:15}, \cite{freyberger/horowitz:15}, \cite{kline/tartari:16}, \cite{laffers:13}, \cite{manski:07,manski:14}, \cite{mogstad/etal:17} and \cite{torgovitsky:16,torgovitsky:18} that similarly propose linear programs to compute identified set in related models. In this proposition, I introduce the following additional quantity $\widetilde{\theta}(Q) \equiv \sum\limits_{w \in \mathcal{W}} a_{\text{num}}(w) \cdot Q(w)~.$

\begin{proposition}\label{prop:bounds}
Suppose that $\mathbf{Q}$ is non-empty and $\theta$ in \eqref{eq:Qtheta} is such that
\begin{align}\label{eq:theta_denominator}
  \sum_{w \in \mathcal{W}} a_{\text{den}}(w) \cdot Q(w) > 0~
 \end{align}
 holds for every $Q \in \mathbf{Q}$. Then the identified set in \eqref{eq:identified_set} can be written as $\Theta = [\theta_l,\theta_u],$ where the endpoints of this interval are solutions to the following two linear programming problems
\begin{align}\label{eq:lin_prog}
  \theta_l = \min_{\gamma, \{ Q(w)\}_{w \in \mathcal{W}}} \widetilde{\theta}(Q)~\text{ and }~\theta_u = \max_{\gamma, \{Q(w)\}_{w \in \mathcal{W}}} \widetilde{\theta}(Q)~,
\end{align}
subject to the following constraints:
\vspace{-.25cm}
 \begin{enumerate}[\enspace (i)]
   \item $\gamma \geq 0$.
   \item $Q(w) \geq 0$ for every $w \in \mathcal{W}$.
   \item $\sum\limits_{w \in \mathcal{W}} Q(w) = \gamma~$.
   \item $\sum\limits_{w \in \mathcal{W}_x} Q(w) = \gamma \cdot \text{Prob}\{Y=y,D=d|Z=z\}$ for every $x = (y,d,z) \in \mathcal{X}~$.
   \item $\sum\limits_{w \in \mathcal{W}_s} Q(w) = 0$.
   \item $\sum\limits_{w \in \mathcal{W}} a_{\text{den}}(w) \cdot  Q(w) = 1~$.
 \end{enumerate}
\end{proposition}

Given the structure of the linear-fractional parameters and the linear restrictions imposed on $Q$, Proposition \ref{prop:bounds} illustrates that the identified set for each parameter is an interval and that the bounds of this interval can be computed by solving two linear programming problems. To do so, the proof first exploits the fact that the identified set can be written as solutions to so-called linear-fractional programs and then results from \cite{charnes/cooper:62} that show how to transform such programs to equivalent linear programs---see also \cite{russell:21} who independently makes a related observation in an alternative treatment effect model.

Computing the identified set using Proposition \ref{prop:bounds} requires two conditions. First, it requires $\mathbf{Q}$ to be non-empty, i.e. the model is correctly specified, to ensure that the identified set is a non-empty interval. If this is not the case, the linear programs automatically terminate. Second, it requires the denominator of the parameter of interest to be positive for every $Q \in \mathbf{Q}$ to ensure that the parameter is well-defined for all feasible model distributions. This condition can easily be verified in practice. To see how, note that the denominator is also a linear-fractional parameter and more specifically a linear parameter. The above proposition can hence be employed to compute the lower bound for this auxiliary parameter to check if it is strictly positive.

\subsection{Statistical Inference}\label{sec:inference}

The above characterization of the identified set assumed that the data moments in \eqref{eq:Qdata} were known with certainty. In practice, we can estimate it using the sample analogue of these moments. I conclude this section by illustrating how to compute confidence intervals for the parameter of interest that account for the resulting statistical uncertainty.

Confidence intervals can be constructed by test inversion. In particular, we can test at level $\alpha \in (0,1)$ the following null hypothesis
\begin{align}\label{eq:null1}
 H_0 : \theta(Q) \equiv \frac{\sum\limits_{w \in \mathcal{W}} a_{\text{num}}(w) \cdot Q(w)}{\sum\limits_{w \in \mathcal{W}} a_{\text{num}}(w) \cdot Q(w)} = \theta_0 \text{ for some } Q \in \mathbf{Q}~,
\end{align}
i.e. the parameter of interest in \eqref{eq:Qtheta} can be equal to some value $\theta_0 \in \mathbf{R}$ given the admissible set of distributions $\mathbf{Q}$. Confidence intervals of level $(1-\alpha)$ for the parameter can then be obtained by collecting the values of $\theta_0$ that are not rejected.

To test this null hypothesis, I propose to use a procedure from \cite{bugni/etal:17}, which they show has attractive theoretical properties that account for the fact that the parameter is partially identified and, as I highlight below, can be computationally attractive in the above setup---see \cite{canay/shaikh:17} for an overview of statistical inference procedures in partially identified setups. Denoting the sample size by $N$, the test is based on a test statistic given by
 \begin{align}\label{eq:TS_N}
   TS(\theta_0) = N \cdot \min_{\{Q(w)\}_{w \in \mathcal{W}}}\sum_{x \in \mathcal{X}}  \left( \hat{P}(x) -  \sum\limits_{\mathcal{W}_x} Q(w)   \right)^2
 \end{align}
subject to the constraints that $Q(w) \geq 0$ for each $w \in \mathcal{W}$, $\sum\limits_{w \in \mathcal{W}} Q(w) = 1$, \eqref{eq:Qass}, and
 \begin{align}\label{eq:Qpara}
   \sum\limits_{w \in \mathcal{W}} a_{\text{num}}(w) \cdot Q(w) = \theta_0 \cdot \sum\limits_{w \in \mathcal{W}} a_{\text{den}}(w) \cdot Q(w)~,
 \end{align}
 where $\hat{P}$ corresponds to the estimated version of $P$ obtained using the empirical distribution of the data, i.e. the test statistic measures the minimum squared distance from zero of the relation between $Q$ and the data in \eqref{eq:Qdata} subject to the various restrictions on $Q$ and that $Q$ can generate $\theta_0$. The test then computes a critical value using the bootstrap. Specifically, for each bootstrap sample $l=1,\ldots,B$, denoting by $\hat{P}_l$ the analogue of $\hat{P}$ computed using the $l$th bootstrap sample, we compute a so-called minimum resampling bootstrap test statistic given by
  \begin{align}\label{eq:TS_MR}
    TS^{MR}_{l}(\theta_0) = \min\{TS_{l}^{DR}(\theta_0),TS_l^{PR}(\theta_0)\}~,
  \end{align}
  where
  \begin{align}\label{eq:TS_DR}
    TS^{DR}_l(\theta_0) = N\sum_{x \in \mathcal{X}}  \left( \hat{P}(x) -  \hat{P}_l(x)   \right)^2
  \end{align}
  is a so-called discard resampling test statistic, and
  \begin{align}\label{eq:TS_PR}
    TS^{PR}_l(\theta_0) = N \cdot \min_{\{Q(w)\}_{w \in \mathcal{W}}}\sum_{x \in \mathcal{X}}  \left( \hat{P}(x) - \hat{P}_l(x) + \frac{1}{\kappa} \left(  \sum\limits_{\mathcal{W}_x} Q(w) - \hat{P}(x) \right)   \right)^2
  \end{align}
  subject to the constraints that $Q(w) \geq 0$ for each $w \in \mathcal{W}$, $\sum\limits_{w \in \mathcal{W}} Q(w) = 1$, \eqref{eq:Qass}, and \eqref{eq:Qpara}, is a so-called penalize resampling test statistic---see \citet[][Section 3]{bugni/etal:17} for intuition behind these two test statistics. Here $\kappa$ is a tuning parameter that satisfies $\kappa \to \infty$ and $\kappa / \sqrt{N} \to 0$ as $N \to \infty$---for example, following \cite{bugni/etal:17}, we can take $\kappa = \sqrt{\text{ln}N}$.

For practical purposes, note that various test statistics in the above procedure can be relatively straightforwardly computed, where that in \eqref{eq:TS_DR} can be directly computed and the optimization problems in \eqref{eq:TS_N} and \eqref{eq:TS_PR} can be computed using a quadratic program given the quadratic objective and linear constraints. The critical value of the test $c(1-\alpha,\theta_0)$ can then be computed by taking the $(1-\alpha)$-quantile of the distribution of bootstrap test statistics in \eqref{eq:TS_MR}, i.e. $\{TS^{MR}_l(\theta_0) : 1 \leq l \leq B\}$, given which the test equals $\phi(\theta_0) = 1\{TS(\theta_0) > c(1-\alpha,\theta_0)\}$, i.e. we reject if the test statistic is larger than the critical value.

\section{Head Start Impact Study}\label{sec:empirical_results}

In this section, I apply the developed tools to the Head Start Impact Study to empirically analyze the average effects of providing an offer to Head Start.

\subsection{Setting}\label{sec:setting}

Head Start is the largest early childhood education program in the United States that provides free preschool education to three- and four-year-old children from disadvantaged households. The Head Start Impact Study (HSIS) was a publicly-funded experiment implemented in Fall 2002 to evaluate Head Start. It was conducted by randomly assigning a sample of three- and four-year-old applicants in certain Head Start preschools to a treatment or control group. Those in the treatment group were provided an offer to Head Start by admitting them in the Head Start preschool to which they had applied, while those in the control were not. However, children in the control group could also receive a Head Start offer if they could be admitted to a preschool not part of the study. See \cite{puma_etal:10} for further details on the program and experiment.

The objective of the analysis is to exploit the experimental variation induced by the HSIS to evaluate the effects of providing an offer to Head Start. I do so by modeling the setting as in Example \ref{ex:hsis} and then evaluating the parameters defined in Section \ref{sec:parameters}. It is useful to highlight that previous analyses of the HSIS, in contrast, have primarily focused on evaluating parameters based on the so-called intention-to-treat (ITT) and instrumental variable (IV) estimands \citep{feller/etal:16, kline/walters:16, puma_etal:10}. The ITT estimands can be defined as
\begin{align}
  \text{ITT}^Y &= E[Y|Z=1] - E[Y|Z=0] ~, \label{eq:ITT_Y}\\
  \text{ITT}^D &= \text{Prob}\{D=p|Z=1\} - \text{Prob}\{D=p|Z=0\}~, \label{eq:ITT_D}
\end{align}
i.e. the difference in outcomes and participation in Head Start between the treatment and control groups, and the IV estimand as
\begin{align}
  \text{IV} = \frac{\text{ITT}^Y}{\text{ITT}^D}~, \label{eq:IV}
\end{align}
i.e. the ratio of the ITT on outcome to that on participation in Head Start. In the following proposition, I illustrate what these estimands allow learning about the effects of a Head Start offer.

\begin{proposition}\label{prop:IV}
 Under Assumption \ref{ass:E} and \eqref{eq:choiceset_hsis}, we have that
 \begin{align}
   \text{ITT}^Y &= E[Y_{C(1)} - Y_{C(0)}]~, \label{eq:ITT_Y_local}\\
   \text{ITT}^D &= \text{Prob}\{D_{C(1)}=p\} - \text{Prob}\{D_{C(0)}=p\}~,  \label{eq:ITT_D_local}\\
   \text{IV} &= E[Y_{C(1)} - Y_{C(1)} | D_{C(1)} = p,~D_{C(0)} \neq p]~. \label{eq:IV_local}
 \end{align}
\end{proposition} 

Proposition \ref{prop:IV} illustrates that these estimands evaluate the effects of the change in choice sets induced by the experiment. They can therefore be interpreted as the policy effects of providing an offer through assignment relative to not through assignment to the treatment and control groups, respectively. However, while those in the treatment group received a Head Start offer and hence $C(1) = C_{+p}$, note that those in the control group could also potentially receive an offer and hence $C(0)$ does not necessarily equal $C_{-p}$. In turn, they do not necessarily answer what are the policy effects of providing an offer to Head Start relative to not more generally. 

Through the defined parameters, the analysis here aims to precisely evaluate such policy effects. As noted in \cite{kline/walters:16}, it is also useful to highlight that providing an offer can open slots at alternative preschools and result in equilibrium changes on the distribution on who has offers to various preschools. In turn, as the proposed model does not account for such general equilibrium effects, the defined parameters may more appropriately be viewed as capturing the policy effects of providing a Head Start offer taking the slots at other schools fixed. Alternatively, they may be viewed as the effects of a marginal policy providing an offer to a single child as such a policy is more likely to keep such slots fixed.

\subsection{Data and Summary Statistics}\label{sec:data}

The analysis uses the data collected by the HSIS. Following \cite{kline/walters:16}, I use a version of the data from the first year of the experiment that pools the two cohorts of three- and four-year-olds together. Moreover, following \cite{feller/etal:16} and \cite{kline/walters:16}, I take the participation decision to be a categorized version of an administratively coded focal care setting variable that evaluated care setting attendance for the entire year. I take the test score outcome to be a discretized version of that used in \cite{kline/walters:16}. They use a summary index of cognitive test scores as measured by the average of the Woodcock Johnson III and the Peabody Picture and Vocabulary Test test scores, where each score is standardized to have mean zero and variance one in the control group for each cohort. Given this summary index, I discretize it using the quantiles of its empirical distribution to take ten support points.\footnote{In unreported results, I compute the bounds in Table \ref{tab:main} using 15 and 20 support points, and find that the numerical results are relatively similar and that the qualitative conclusions remain unchanged.} See Appendix \ref{S-sec:empirical_add} for further details on how the analysis sample was constructed.

Table \ref{tab:summ} provides some summary statistics for the analysis sample. Panel (a) reports the participation shares across the different types of care settings by treatment and control groups. A large proportion (82.5$\%$) of the treatment group participated in Head Start, which suggests that many prefer Head Start over their alternative options. Comparing to those in Head Start in the control group suggests that many control children may not have had a Head Start offer. However, given that this proportion is not zero, this suggests that at least some proportion (14$\%$) in fact received an offer even when assigned to the control group.

\begin{table}[!t]
\begin{center}
\caption{Summary statistics}
\label{tab:summ}
\scalebox{0.8}{
\def\arraystretch{0.9}
\makebox[\textwidth]{\begin{threeparttable}
\input{HSsumm.tex}
\begin{tablenotes}[flushleft]
\setlength\labelsep{0pt}
\item \textit{Notes}: Parentheses report standard errors clustered at Head Start center level.
\end{tablenotes}
\end{threeparttable} }}
\end{center}
\end{table}

Table \ref{tab:summ} Panel (b) reports estimates for the estimands in \eqref{eq:ITT_Y}-\eqref{eq:IV}. As $\text{ITT}^Y$ does not require the test scores to be discrete, I also report results under the raw undiscretized version of the test scores---the results under both versions are relatively similar. Since some children in the control group received a Head Start offer, these estimands, as shown in Proposition  \ref{prop:IV}, generally evaluate the effect of providing a Head Start offer through assignment to treatment group relative to the control group. The IV estimate reveals such a policy results in a positive effect of 0.249 standard deviation points for the individuals who are affected by it. 

\subsection{Effects of a Head Start Offer}\label{sec:estimates}

Table \ref{tab:main} reports estimated identified sets and 90$\%$ confidence intervals for the parameters evaluating the average effect of a Head Start offer in \eqref{eq:ATE_Y}-\eqref{eq:ATOP} when taking $d$ to be equal to $p$, i.e. Head Start. The different columns correspond to different combinations of assumptions, which we motivate and describe below. As in the rest of the analysis below, the identified sets are estimated by applying the linear programming procedure from Proposition \ref{prop:bounds} to the empirical distribution of the data, and the confidence intervals are constructed as in Section \ref{sec:inference} where the bootstrap samples are drawn at the level of Head Start centers.

\begin{table}[!t]
\begin{center}
\caption{Average effects of a Head Start offer}
\label{tab:main}
\scalebox{0.8}{
\def\arraystretch{0.9}
\makebox[\textwidth]{\begin{threeparttable}
\input{HSmain.tex}
\begin{tablenotes}[flushleft]
\setlength\labelsep{0pt}
\item \textit{Notes}: For each parameter, the inner (outer) panel reports the lower and upper bounds (endpoints of the confidence interval), respectively. Lower and upper bounds are not repeated when they coincide. Assumptions \ref{ass:MTR}($\lambda$) and \ref{ass:Roy}($\lambda$) refer to Assumptions \ref{ass:MTR} and \ref{ass:Roy} imposed on $\lambda$ proportion of children as in \eqref{eq:mixture}.
\end{tablenotes}
\end{threeparttable} }}
\end{center}
\end{table}

Under the baseline specification in Column (1), $\text{ATE}^D(C_{+p},C_{-p})$ reveals that providing a Head Start offer affects a large percentage of children (85.5$\%$) who take up the offer and participate in Head Start. Note that this parameter is point identified as $C_{+p} = C(1)$ given \eqref{eq:choiceset_hsis} and is hence equal to the proportion participating in Head Start in the treatment group from Table \ref{tab:summ} Panel (a). Turning to the effects on test scores, $\text{ATE}^Y(C_{+p},C_{-p})$ and $\text{ATOP}(C_{+p},C_{-p})$ reveal that the upper bound can be potentially large and greater than the IV estimand, highlighting that large test scores are consistent with the data. However, the lower bound is smaller than zero and hence the data alone does not allow us to conclude whether we have positive or negative.

Intuitively, this arises because the baseline specification leaves the relationship between the test scores and other model variables completely unrestricted. To this end, I consider the following additional assumptions restricting the relation that help is reaching more informative conclusions, and which can be written as \eqref{eq:Qass} as shown in Appendix \ref{S-sec:rewriting_ass}.

\begin{assumption2}{NC}\label{ass:NC}
 (Nested Choice) $d(U,\{n,p\}) = p$ if and only if $d(U,\{n,a\}) = a$~.
\end{assumption2}

\begin{assumption2}{MTR}\label{ass:MTR}
(Monotone Treatment Response) $Y(p) \geq Y(n)$ and $Y(p) \geq Y(a)~$.
\end{assumption2}

\begin{assumption2}{Roy}\label{ass:Roy}
For each $d,d' \in \mathcal{D}$, if $Y(d') > Y(d)$ then $d(U,\{d,d'\}) = d'$~.
\end{assumption2}

\noindent Assumption \ref{ass:NC} imposes parents to prefer Head Start to home care if and only if they also prefer alternative preschools to home care. It is motivated by the idea that parents may have preferences where they prefer to enroll their child in any preschool, Head Start or an alternative one, to home care or vice versa. Assumption \ref{ass:MTR}, based on \cite{manski:97b}, imposes test scores under Head Start to be at least as good as those under alternative preschools and home care. It is motivated by the notion that since Head Start is considered to be a high quality care setting and that families in Head Start are primarily disadvantaged, it may be that Head Start is a better option than home care and alternative preschools that these families could afford. Finally, Assumption \ref{ass:Roy} imposes parents to prefer enrolling their child in the care setting with the higher test score, i.e. it takes preferences to be based solely on maximizing outcomes in the spirit of the Roy selection model \citep{heckman/honore:90, mourifie/etal:15}.

While Assumption \ref{ass:NC} is reasonable, Assumptions \ref{ass:MTR} and \ref{ass:Roy} may seem potentially strong to impose for every child in the population. To this end, I also consider weaker versions of them, where I assume that they hold for only some pre-specified proportion of children. More formally, let the primitive distribution of the model be given by
\begin{align}\label{eq:mixture}
 Q(w) = \lambda \cdot H_1(w) + (1- \lambda) \cdot H_0(w)
\end{align}
for each $w \in \mathcal{W}$, where $\lambda \in [0,1]$ corresponds to the known proportion of children for which either Assumptions \ref{ass:MTR} or \ref{ass:Roy} hold, and $H_1$ and $H_0$ correspond to the underlying distribution for the children for whom these assumptions hold and may not hold, respectively---i.e., the restrictions imposed by these assumptions of the form \eqref{eq:Qass} are imposed only on $H_1$ and not on $H_0$, which are otherwise identical in term of mass assigned to preferences and choice sets. Indeed, if $\lambda$ equals one, these weaker versions are equivalent to Assumptions \ref{ass:MTR} or \ref{ass:Roy}. However, by taking this proportion to be smaller than one, we can make these assumptions flexibly weaker and analyze sensitivity of the empirical conclusions to the assumptions---similar in spirit to the breakdown point analysis considered in \cite{horowitz/manski:95}, \cite{kline/santos:13} and \cite{masten/poirier:17}. In Appendix \ref{S-sec:modifiedLP}, I illustrate how Proposition \ref{prop:bounds} can be straightforwardly modified to account for these weaker versions of Assumptions \ref{ass:MTR} or \ref{ass:Roy}.

The results are presented in Columns (2)-(7) of Table \ref{tab:main}. From Columns (3) and (6), we can observe that under Assumptions \ref{ass:MTR} or \ref{ass:Roy}, the lower bound of $\text{ATOP}(C_{+p},C_{-p})$ significantly increases and allows concluding that the offer can have a large positive benefit to children who take it up. In particular, their test scores on average can increase between 0.206 and 0.497 standard deviation points. Moreover, from Columns (4)-(5) and (7)-(8), these conclusions can continue to hold under more flexible specifications that impose these additional assumptions not on all children but only on some given proportion, namely $80 \%$ and $90 \%$ in this case.

\subsection{Heterogeneity Based on an Alternative Preschool Offer}\label{sec:alternative}

The above results reveal that a Head Start offer affects a large number of children who take up the offer, and can have a positive effect on their subsequent test scores. To better understand these effects and their driving factors, I next analyze heterogeneity based on whether a child had an offer or not to an alternative (non-Head Start) preschool, i.e. the parameters in \eqref{eq:ATE_D_a+}-\eqref{eq:ATOP_a-} taking $d' = a$.

Table \ref{tab:alternative} reports estimated identified sets and 90$\%$ confidence intervals for these parameters. Observe that parameters evaluating the effects for the group who do not have an alternative preschool offer are not well-defined under all the specifications. This is because the probability of this subgroup, corresponding to the conditioning event of these parameters, can potentially be zero given the data and specification. In turn, since this implies that \eqref{eq:theta_denominator} is not satisfied, these parameters are not well-defined.\footnote{In this case, the confidence intervals equal the logical bounds under these specifications as the numerator also equals zeros when the denominator does for these parameters and hence all values of $\theta_0$ permitted under the specification can satisfy \eqref{eq:Qpara}.} However, when Assumption \ref{ass:NC} is imposed, this is no longer the case as the probability of the conditioning event is bounded away from zero.

\begin{table}[!t]
\begin{center}
\caption{Heterogeneity in the availability of an alternative preschool offer}
\label{tab:alternative}
\scalebox{0.8}{
\def\arraystretch{0.9}
\makebox[\textwidth]{\begin{threeparttable}
\input{HShetero.tex}
\begin{tablenotes}[flushleft]
\setlength\labelsep{0pt}
\item \textit{Notes}: For each parameter, the inner (outer) panel reports the lower and upper bounds (endpoints of the confidence interval), respectively. `-' indicates \eqref{eq:theta_denominator} is not satisfied. Specifications correspond to columns from Table \ref{tab:main}.
\end{tablenotes}
\end{threeparttable} }}
\end{center}
\end{table}

The results reveal that providing a Head Start offer affects more children who don't have an alternative preschool offer than those who do. In particular, Column (2) reveals that between 57.5$\%$ and 82.2$\%$ of children with an alternative preschool offer take up the Head Start offer in comparison to between 82.8$\%$ and 100$\%$ of those without an alternative preschool offer. Moreover, under Assumptions \ref{ass:MTR} or \ref{ass:Roy}, the results reveal that providing an offer can potentially have positive test score gains for only those who don't have an alternative preschool offer and not for those who do. These results together highlight that the effects of providing an offer are primarily driven by the subgroup of children who do not have an alternative preschool offer and hence any outside preschool option.

\subsection{Cost-Benefit Analysis of a Head Start Offer}\label{sec:cost_benefit}

The motivation to study the above effects of a Head Start offer was to capture the potential costs and benefits of a policy providing an offer. I conclude the empirical analysis by using the various effects to perform a cost-benefit analysis of such a policy, similar in spirit to that performed by \cite{kline/walters:16} using the local effects. The analysis can be done by attaching an income value to the test score gains to evaluate the benefits as well as one to the take up of Head Start offer to evaluate the costs. To this end, let $\rho$ denote the income value for the test score gains and $\phi_p$ denote the cost of a student participating in Head Start. Moreover, as emphasized in \cite{kline/walters:16}, it is also important when measuring the costs to account for the potential cost savings associated with the movement of children out of alternative preschools, as these preschools are often publicly funded as well. To this end, let $\phi_a$ analogously denote the cost of participating in an alternative preschool.

Using these values, the average benefits of providing a Head Start offer relative to not can be given by the average effect of test scores times the income value, i.e.
\begin{align*}
\text{AB} = \rho \cdot \text{ATE}^Y(C_+,C_-)~.
\end{align*}
Similarly, the average costs can be given by the proportion who take up the Head Start offer times the cost of participation along with subtracting the costs associated with those who were induced into Head Start from an alternative preschool, i.e.
\begin{align*}
 \text{AC} = \phi_p \cdot \text{ATE}^D(C_+,C_-) - \phi_a \cdot \text{Prob}\{D_{C_+} = p, D_{C-} = a \}~.
\end{align*}
Taking the difference of the benefit and cost parameters, we can then define the average surplus
\begin{align*}
 \text{AS} = \text{AB} - \text{AC}~,
\end{align*}
which can be used to perform a cost-benefit of a policy providing an offer. Given values of $\rho$, $\phi_p$ and $\phi_a$, these parameters, similar to \eqref{eq:ATE_D}-\eqref{eq:ATOP_a-}, can be rewritten as \eqref{eq:Qtheta}. In turn, the linear programming procedure can be continued to be applied to characterize the identified set for these parameters. Following \cite{kline/walters:16}, I take $\rho = \kappa \cdot E$, where $E$ corresponding to the average present discounted value of lifetime earnings for Head Start applicants is taken to be $\$$34,339 and $\kappa$ corresponding to the relationship between earnings and a one standard deviation increase in test scores is taken to be either 0.1 or 0.3, which is based on the lower and higher end of the values for this relation noted in the literature \citep[][Table A.IV]{kline/walters:16}. Moreover, following  \cite{kline/walters:16}, I take $\phi_p = \$ 8,000 $ and $\phi_a = 0.75 \cdot \phi_p = \$ 6,000$.

\begin{table}[!t]
\begin{center}
\caption{Cost-benefit analysis of providing a Head Start offer}
\label{tab:cost_benefit}
\scalebox{0.8}{
\def\arraystretch{0.9}
\makebox[\textwidth]{\begin{threeparttable}
\input{HScostbenefit.tex}
\begin{tablenotes}[flushleft]
\setlength\labelsep{0pt}
\item \textit{Notes}: For each parameter, the inner (outer) panel reports the lower and upper bounds (endpoints of the confidence interval), respectively. Specifications correspond to columns from Table \ref{tab:main}.
\end{tablenotes}
\end{threeparttable} }}
\end{center}
\end{table}

Table \ref{tab:cost_benefit} reports estimated identified sets and 90$\%$ confidence intervals for these parameters for a subset of the specifications from Table \ref{tab:main} that impose restrictions on test scores. The results reveal that the net costs of providing an offer is relatively low in comparison to the absolute value of participating in Head Start. Similar to that highlighted by \cite{kline/walters:16}, this arises due to relatively high cost savings from children who enroll out of alternative preschools when provided with a Head Start offer. Taking the benefits into account, the upper bounds of the average surplus reveals that benefits can significantly outweigh costs between the $\$$9,727 and $\$$37,873 per child depending on the relation between test scores and earnings. However, whether the lower bound is positive such that the surplus is in fact positive depends on the strength of the assumptions and the relation between test scores and earnings. Specifically, when Assumptions \ref{ass:MTR} or \ref{ass:Roy} are imposed on all children in Column (4) and (7), the surplus can be positive. But, when these assumptions are imposed only on some proportion of the children in the remaining columns, it can be positive only when $\kappa = 0.3$, i.e. the relation between test scores gains and earnings is based on those towards the higher end of that noted in the literature \citep[e.g.,][]{heckman/etal:10}.

\section{Conclusion}\label{sec:conclusion}

I propose a novel treatment selection model with unobserved heterogeneity in both choice sets and preferences to evaluate the average effects of a program offer. I show how to exploit the structure of the model to define parameters capturing these effects and then computationally characterize their identified sets under instrumental variable variation in choice sets. I illustrate these tools by analyzing the effects of providing a Head Start offer using data from the Head Start Impact Study.


\newpage
\bibliography{references.bib}

\end{document}


\author{
Vishal Kamat \\
Toulouse School of Economics\\
University of Toulouse Capitole\\
\url{vishal.kamat@tse-fr.eu} 
}

\title{Supplementary Appendix to ``Identifying the Effects of a Program Offer with an Application to Head Start''}

\maketitle

\begin{abstract}
This document provides proofs and additional details pertinent to the author's paper ``Identifying the Effects of a Program Offer with an Application to Head Start." Section \ref{sec:modifiedLP} describes the modified procedure used to compute bounds under weaker versions of the imposed assumptions. Section \ref{sec:empirical_add} provides additional details pertinent to the empirical analysis. Section \ref{sec:proofs} provides the proofs and other mathematical derivations.
\end{abstract}

\thispagestyle{empty}

\newpage
\setcounter{page}{1}

\appendix
\renewcommand{\theequation}{S.\arabic{equation}}
\renewcommand{\thetable}{S.\arabic{table}}
\renewcommand{\thesection}{S.\arabic{section}}

\section{Modified Linear Program}\label{sec:modifiedLP}

In this section, I describe how the linear programming procedure from Proposition \ref{O-prop:bounds} can be modified to allow for weaker versions of various assumptions in the form of \eqref{O-eq:mixture}. The arguments behind this modified procedure are similar to those of Proposition \ref{O-prop:bounds} presented in Section \ref{O-sec:computation} with a few differences. Instead of the unknown quantity being only $Q$, there are two additional auxiliary unknown probability mass functions $H_0$ and $H_1$ with support similarly contained in $\mathcal{W}$, i.e. $H_0: \mathcal{W} \to [0,1]$ and  $H_1: \mathcal{W} \to [0,1]$ such that $\sum\limits_{w \in \mathcal{W}} H_0(w) = 1$ and $\sum\limits_{w \in \mathcal{W}} H_1(w) = 1$.

Given these additional unknown quantities, we need to introduce the additional restrictions over them which then indirectly introduce restrictions on $Q$ through their relationship to it. To this end, as $H_1$ and $H_0$ are identical in terms of mass associated to preferences and choice sets, we have
\begin{align}\label{eq:Hcomb_SA}
 \sum_{(y(d) : d \in \mathcal{D}) \in \mathcal{Y}^{|\mathcal{D}|}} H_0(w) = \sum_{(y(d) : d \in \mathcal{D}) \in \mathcal{Y}^{|\mathcal{D}|}} H_1(w)
\end{align}
for each $(u,\{c(z) : z \in \mathcal{Z}\}) \in \mathcal{U}\times \mathcal{C}^{|\mathcal{Z}|}$, where $w = (\{y(d) : d \in \mathcal{D} \},u,\{c(z) :  \in \mathcal{Z}\})$. Moreover, let $s_1$ denote the assumptions we want to weaken, for example Assumptions \ref{O-ass:MTR} or \ref{O-ass:Roy} in the application, which are imposed only on $H_1$ and not on $H_0$, i.e.
\begin{align}\label{eq:Hrest_SA}
\sum_{w \in \mathcal{W}_{s_1}} H_1(w) = 0~,
\end{align}
where $\mathcal{W}_{s_1}$ is a known subset of $\mathcal{W}$. Given these restrictions on the auxiliary probability mass functions, the relationship between them and $Q$ then indirectly impose restrictions on $Q$, which recall is given by
\begin{align}\label{eq:Qmix_SA}
 Q(w) = \lambda \cdot H_1(w) + (1- \lambda) \cdot H_0(w)
\end{align}
for each $w \in \mathcal{W}$, where $\lambda \in [0,1]$ is a pre-specified known proportion. To summarize, what we know about $Q$ from these restrictions imposed through the auxiliary distributions can then be captured by the following set
\begin{align}\label{eq:Qfeasbile_aux}
 \mathbf{Q}_{\lambda,\text{aux}} = \{ Q \in \mathbf{Q}_{\mathcal{W}} : Q \text{ satisfies \eqref{eq:Qmix_SA}; $H_0,H_1 \in \mathbf{Q}_{\mathcal{W}}$ satisfy \eqref{eq:Hcomb_SA}-\eqref{eq:Hrest_SA}} \}~,
\end{align}
where recall that $\mathbf{Q}_{\mathcal{W}}$ is the set of all probability mass functions on the sample space $\mathcal{W}$.

Using the notation from Section \ref{O-sec:identifiedset},  the identified set for a pre-specified parameter $\theta(Q)$ and for a pre-specified $\lambda$ can then be written as follows
\begin{align}\label{eq:identified_set_lambda}
 \Theta_{\lambda} = \{ \theta_0 \in \mathbf{R} : \theta(Q) = \theta_0 \text{ for some } Q \in \mathbf{Q}_{\lambda}  \}~,
\end{align}
where $\mathbf{Q}_{\lambda} = \mathbf{Q} \cap \mathbf{Q}_{\lambda,\text{aux}}$ is the intersection of the set of feasible distributions determined by restrictions directly imposed on $Q$ by the data and assumptions as described in Section \ref{O-sec:identifiedset} and indirectly imposed on $Q$ through the restrictions imposed on the auxiliary distributions as described above.

Given the linear-fractional structure of the parameter and the linear structure of all the restrictions, we can use the same arguments as those in Proposition \ref{O-prop:bounds} to show that a linear program can be used to compute the identified set in this case. Indeed, the only difference between the linear program in this proposition and that in Proposition \ref{O-prop:bounds} is the introduction of additional variables and linear restrictions on these variables.

\begin{proposition}\label{prop:LPsensitivity}
Suppose that $\mathbf{Q}_{\lambda}$ is non-empty and the parameter characterized by \eqref{O-eq:Qtheta} is such that
\begin{align}\label{eq:theta_denominator_lambda}
  \sum_{w \in \mathcal{W}} a_{\text{den}}(w) \cdot Q(w) > 0~
 \end{align}
 holds for every $Q \in \mathbf{Q}_{\lambda}$. Then the identified set in \eqref{eq:identified_set_lambda} can be written as $\Theta_{\lambda} = [\theta_l,\theta_u]$, where the lower and upper bounds of this interval are solutions to the following two linear programming problems
\begin{align}\label{eq:lin_prog_lambda}
  \theta_l = \min_{\gamma, \{ Q(w), H_0(w),H_1(w)\}_{w \in \mathcal{W}}} \widetilde{\theta}(Q)~\text{ and }~\theta_u = \max_{\gamma, \{Q(w), H_0(w),H_1(w)\}_{w \in \mathcal{W}}} \widetilde{\theta}(Q)~,
\end{align}
subject to the following constraints:
\vspace{-.25cm}
 \begin{enumerate}[\enspace (i)]
   \item $\gamma \geq 0~$.
   \item $Q(w),~H_0(w),~H_1(w) \geq 0$ for every $w \in \mathcal{W}~$.
   \item $\sum\limits_{w \in \mathcal{W}} Q(w) = \gamma$, $\sum\limits_{w \in \mathcal{W}} H_0(w) = \gamma$, $\sum\limits_{w \in \mathcal{W}} H_1(w) = \gamma$~.
   \item $\sum\limits_{w \in \mathcal{W}_x} Q(w) = \gamma \cdot \text{Prob}\{Y=y,D=d|Z=z\}$ for every $x = (y,d,z) \in \mathcal{X}~$.
   \item $\sum\limits_{w \in \mathcal{W}_s} Q(w) = 0$, $\sum\limits_{w \in \mathcal{W}_{s_1}} H_1(w) = 0$~.
   \item $\sum\limits_{w \in \mathcal{W}} a_{\text{den}}(w) \cdot Q(w) = 1~$.
   \item $Q(w) = \lambda \cdot H_1(w) + (1- \lambda) \cdot H_0(w)$ for every $w \in \mathcal{W}~$.
   \item $\sum\limits_{(y(d) : d \in \mathcal{D}) \in \mathcal{Y}^{|\mathcal{D}|}} H_0(w) = \sum\limits_{(y(d) : d \in \mathcal{D}) \in \mathcal{Y}^{|\mathcal{D}|}} H_1(w)$ for every $(u,\{c(z) : z \in \mathcal{Z}\}) \in \mathcal{U}\times \mathcal{C}^{|\mathcal{Z}|}$~.
 \end{enumerate}
\end{proposition}

\section{Additional Details for Empirical Analysis}\label{sec:empirical_add}

\subsection{Details on Discretizing Test Scores}\label{sec:discretization}

In this section, I outline how I obtain the discrete test score outcome of interest used in the empirical results presented in Section \ref{O-sec:empirical_results}. I use the empirical distribution of the observed undiscretized test scores to discretize the test scores to take values in $\mathcal{Y} = \{y_1, \ldots, y_M\}$. Specifically, for a pre-specified choice of $M$, I take each point in $\mathcal{Y}$ to be given by
\begin{align*}
 y_m = \frac{y^{*}_m + y^{*}_{m-1}}{2}~,
\end{align*}
where $y^*_m$ denotes the $m/M$-quantile of the empirical distribution of the observed undiscretized test score $Y^*$, i.e.  each point is determined by the midpoint of two specific adjacent quantiles of the empirical distribution of the observed undiscretized test scores. Then, each observed undiscretized test score $Y^*$ is transformed as follows to its corresponding discretized version
\begin{align*}
 Y =
 \begin{cases}
  y_m \text{ if } Y^{*} \in \left[ y^{*}_{m-1}, y^{*}_{m} \right) \text{ for } m \in \{1, \ldots, M-1 \}~, \\
  y_M \text{ if } Y^{*} \in \left[ y^{*}_{M-1}, y^{*}_{M} \right]~,
 \end{cases}
\end{align*}
which is the version of the test score used in the analysis. For the empirical results, I used ten support points for the test scores, i.e. $M=10$, which ensured that all the computational problems were generally tractable. In unreported results, I also computed estimated bounds for the parameters in Table \ref{O-tab:main} with $M=15$ and $M=20$. I find that the numerical results are relatively similar and, moreover, that the qualitative conclusions based on them remain unchanged. 

\subsection{Data and Variable Construction}\label{sec:data}

The raw data I use from the Head Start Impact Study (HSIS) is restricted, but access can be acquired by submitting applications to Research Connections at \url{http://www.researchconnections.org/childcare/resources/19525}. In this section, I briefly describe how the raw data was transformed to the final analysis sample, which closely followed the publicly available code used to construct the analysis sample in \cite{kline/walters:16}. In particular, I first merged all the various data files provided by Research Connections and dropped observations with missing Head Start preschool IDs, where this preschool corresponded to that from which the child was sampled. I then classified the participation decision into the three categories used in the paper using the focal care arrangement variable provided by the HSIS data set. Finally, all observations where any of the variables used in the analysis were missing were dropped.

\section{Proofs of Propositions and Additional Derivations}\label{sec:proofs}

\subsection{Proof of Proposition \ref{O-prop:bounds}}\label{sec:proof_bounds}

Note that $\mathbf{Q}_{\mathcal{W}}$ is compact and convex and that $\mathbf{Q}$ is a subset of $\mathbf{Q}_{\mathcal{W}}$ obtained by placing the linear constraints in \eqref{O-eq:Qdata} and \eqref{O-eq:Qass}. This in turn implies that $\mathbf{Q}$ is a compact and convex set as well.

Next, it follows from \eqref{O-eq:Qtheta} that $\theta(Q)$ is a linear-fractional function of $Q$ where the denominator is required to be positive, i.e. \eqref{O-eq:theta_denominator} holds, for every $Q \in \mathbf{Q}$. Along with $\mathbf{Q}$ being a compact and convex set, this in turn implies that $\theta(\mathbf{Q})$ is a compact and convex set in $\mathbf{R}$. More specifically, it follows that the identified set in \eqref{O-eq:identified_set} can be written as a closed interval, where the lower bound and upper bound are given by
\begin{align}\label{eq:LP}
  \theta_l = \min_{ Q \in \mathbf{Q}} \theta(Q)~\text{ and }~\theta_u = \max_{ Q \in \mathbf{Q}} \theta(Q)~.
\end{align}

In order to complete the proof, note that the optimization problems in \eqref{eq:LP} have linear-fractional objectives due to the structure of the parameter in \eqref{O-eq:Qtheta} and a finite number of linear constraints as guaranteed by the restrictions in \eqref{O-eq:Qdata} and \eqref{O-eq:Qass}. Such optimization problems are commonly referred to as linear-fractional programs. For such programs, \cite{charnes/cooper:62} show that if the feasible set of the program is non-empty and bounded, and if the denominator of the linear-fractional objective is strictly positive for all values in the feasible set, then the linear-fractional program can be transformed to an equivalent linear program---see \citet[][Section 4.3.2]{boyd/vand:04} for a textbook exposition of this result.

For the linear-fractional programs stated in \eqref{eq:LP}, both these conditions are satisfied. Since $\mathbf{Q}$ is non-empty and bounded, we have that the feasible sets of the programs given by $\mathbf{Q}$ are indeed non-empty and bounded. Further, since \eqref{O-eq:theta_denominator} holds for all $Q \in \mathbf{Q}$, we also have that the denominator of the objectives are strictly positive for all values in the feasible set. In turn, the result from \cite{charnes/cooper:62} can be invoked to transform the linear-fractional programs in \eqref{eq:LP} to equivalent linear programs which are given by those in \eqref{O-eq:lin_prog}. Specifically, the equivalent linear programs are obtained by introducing the following transformation
\begin{align}\label{eq:CCtransform}
\widetilde{Q}(w) =  \gamma \cdot Q(w)~\text{ where }~\gamma = \ddfrac{1}{\sum_{w \in \mathcal{W}} a_{\text{den}}(w) \cdot Q(w)}~,
\end{align}
which is well-defined given that \eqref{O-eq:theta_denominator} holds for every $Q \in \mathbf{Q}$, and by introducing the additional constraint that $\gamma \geq 0$. Then, when the constraints and the objectives stated in terms of $Q$ for the linear-fractional programs in \eqref{eq:LP} and the relationship between $Q$ and $\gamma$ in \eqref{eq:CCtransform} are rewritten in terms of $\gamma$ and $\widetilde{Q}$, we obtain the constraints and the objective of the linear programs stated in \eqref{O-eq:lin_prog}, which concludes the proof.

\subsection{Proof of Proposition \ref{O-prop:IV}}\label{sec:local}

In order to see \eqref{O-eq:ITT_Y_local}, note that
\begin{align*}
  \text{ITT}^Y &= E[Y|Z=1] - E[Y|Z=0] \\
  &= E[Y_{C(1)}|Z=1] - E[Y_{C(0)}|Z=0] \\
  &= E[Y_{C(1)} - Y_{C(0)}]~,
\end{align*}
where the first equality recalls the definition of $\text{ITT}^Y$ from \eqref{O-eq:ITT_Y}, the second equality follows from \eqref{O-eq:Y_equation}, \eqref{O-eq:selection_equation} and \eqref{O-eq:C_equation}, and the definition that $Y_c = Y(D_c)$ for each non-empty $c \subseteq \mathcal{D}$, and the final equality from Assumption \ref{O-ass:E}. In an analogous manner, by replacing $Y$ with $1\{D=p\}$ we can show \eqref{O-eq:ITT_D_local}.

In order to see \eqref{O-eq:IV_local}, note first that
\begin{align}
  \text{ITT}^Y &= E[Y_{C(1)} - Y_{C(0)}] \nonumber \\
  &= E[Y_{C(1)} - Y_{C(0)}|D_{C(1)} \neq D_{C(0)}] \text{Prob}\{D_{C(1)} \neq D_{C(0)}\} \nonumber \\
  &=  E[Y_{C(1)} - Y_{C(0)}|D_{C(1)} = p,~ D_{C(0)} \neq p] \text{Prob}\{D_{C(1)} = p,~ D_{C(0)} \neq p\}~, \label{eq:ITT_Y_proof}
\end{align}
where the first equality follows from \eqref{O-eq:ITT_Y_local}, the second equality follows from the fact that $Y_{C(1)} \equiv Y(D_{C(1)}) = Y(D_{C(0)}) \equiv Y_{C(0)}$ if $D_{C(1)} = D_{C(0)}$, and the third equality follows from \eqref{O-eq:choiceset_hsis} given which $1\{D_{C(1)} \neq D_{C(0)}\} = 1\{D_{C(1)}=p,~D_{C(0)} \neq p\}$. In an analogous manner, we can show
\begin{align}\label{eq:ITT_D_proof}
  \text{ITT}^D = \text{Prob}\{D_{C(1)} = p,~ D_{C(0)} \neq p\}~.
\end{align}
Dividing \eqref{eq:ITT_Y_proof} by \eqref{eq:ITT_D_proof} then implies \eqref{O-eq:IV_local}, which completes the proof.

\subsection{Rewriting Parameters in Section \ref{O-sec:parameters} in terms of \eqref{O-eq:Qtheta} for Example \ref{O-ex:hsis}}\label{sec:rewriting_theta}

Similar to how the parameter \eqref{O-eq:ATE_D} in the context of Example \ref{O-ex:hsis} when $d=p$ can be written as \eqref{O-eq:Qtheta} as illustrated in \eqref{O-eq:ATE_D_Q}, I show how to do for the remaining parameters in Section \ref{O-sec:parameters} in the context of Example \ref{O-ex:hsis}, which are empirically analyzed in Section \ref{O-sec:empirical_results}. To this end, as $C_{+p} = C(1)$ and $C_{-p} = C(1) \setminus \{p\}$ given \eqref{O-eq:C_equation} and \eqref{O-eq:choiceset_hsis}, the parameter \eqref{O-eq:ATE_Y} when $d=p$ can be written as
\begin{align*}
  \text{ATE}^Y(C_{+p},C_{-p})(Q) =  \sum_{w \in \mathcal{W}}  \cdot [y(d( u,c(1)  ) ) - y(d(u,c(1)\setminus \{p\} ) )] \cdot Q(w)~,
\end{align*}
while the parameter in \eqref{O-eq:ATOP} can be written as the ratio of that above and that in \eqref{O-eq:ATE_D_Q}. Similarly, the parameters in \eqref{O-eq:ATE_D_a+} and \eqref{O-eq:ATOP_a+} can be respectively written as
\begin{align*}
  \text{ATE}^D_{+a}(C_{+p},C_{-p})(Q) &=  \frac{\sum\limits_{w \in \mathcal{W}_{+a}^p}  Q(w)}{  \sum\limits_{w \in \mathcal{W}_{+a}}  Q(w) } ~, \\
\text{ATOP}_{+a}(C_{+p},C_{-p})(Q) &= \frac{\sum\limits_{w \in \mathcal{W}_{+a}} [y( d(u,c(1) ) ) - y( d( u, c(1)\setminus \{p\}  )  )] \cdot Q(w)}{ \sum\limits_{w \in \mathcal{W}^p_{+a}} Q(w)  }~, 
\end{align*} 
where $\mathcal{W}^p_{+a} = \{w \in \mathcal{W} : d(u, c(1)) = p,~ a \in c(z) \}$ and $\mathcal{W}_{+a} = \{w \in \mathcal{W} : a \in c(1)\}$, and the parameters in \eqref{O-eq:ATE_D_a-} and \eqref{O-eq:ATOP_a-} can be respectively written as 
\begin{align*}
 \text{ATE}^D_{-a}(C_{+p},C_{-p})(Q) &=  \frac{\sum\limits_{w \in \mathcal{W}_{-a}^p}  Q(w)}{  \sum\limits_{w \in \mathcal{W}_{-a}}  Q(w) } ~,  \\
\text{ATOP}_{-a}(C_{+p},C_{-p})(Q) &= \frac{\sum\limits_{w \in \mathcal{W}_{-a}}  [y( d(u,c(1) ) ) - y( d( u, c(1) \setminus \{p\}  )  )] \cdot Q(w)}{ \sum\limits_{w \in \mathcal{W}^h_{-a}} Q(w)  }~,
\end{align*}
where $\mathcal{W}^p_{-a} = \{w \in \mathcal{W} : d(u, c(1)) = p,~ a \notin c(1) \}$ and $\mathcal{W}_{-a} = \{w \in \mathcal{W} : a \notin c(1)\}$.

\subsection{Rewriting Assumptions in Section \ref{O-sec:estimates} in terms of \eqref{O-eq:Qass}}\label{sec:rewriting_ass}

Similar to \eqref{O-eq:choiceset_hsis} in Section \ref{O-sec:identifiedset}, I illustrate in this section how the various additional assumptions used in the empirical analysis in Section \ref{O-sec:estimates} can be re-written as restrictions on $Q$ in the form of \eqref{O-eq:Qass}. For example, let $\mathcal{W}_s$ be the subset of $\mathcal{W}$ using which \eqref{O-eq:Qass} is imposed in the absence of Assumption \ref{O-ass:NC}. In turn, if Assumption \ref{O-ass:NC} is imposed, it is straightforward to see that one can do so by redefining $\mathcal{W}_s$ in \eqref{O-eq:Qass} by $\mathcal{W}_s \equiv \mathcal{W}_s \cup \mathcal{W}_{\text{NC}}$, where
\begin{align*}
  \mathcal{W}_{\text{NC}} = \{ w \in \mathcal{W} : d_{u,\{n,p\}} = p,~ d_{u,\{n,a\}} = n \text{~or~}  d_{u,\{n,p\}} = n,~ d_{u,\{n,a\}} = a\}~.
\end{align*}
Similarly, if Assumption \ref{O-ass:MTR} is imposed, then $\mathcal{W}_s$ can redefined by $\mathcal{W}_s \equiv \mathcal{W}_s \cup \mathcal{W}_{\text{MTR}}$, where
\begin{align*}
  \mathcal{W}_{\text{MTR}} = \{ w \in \mathcal{W} : y(n) > y(p) \text{ or } y(a) > y(p) \}~.
\end{align*}
In order to see Assumption \ref{O-ass:Roy} can be imposed, note first this assumption can be equivalently stated as $d(U,\{d,d'\}) = d~\implies~Y(d') \leq Y(d)$ for each $d,d' \in \mathcal{D}$. Then, it follows that if Assumption \ref{O-ass:Roy} is imposed,  then $\mathcal{W}_s$ can redefined by $\mathcal{W}_s \equiv \mathcal{W}_s \cup \mathcal{W}_{\text{Roy}}$, where
\begin{align*}
 \mathcal{W}_{\text{Roy},\{d,d'\}} = \bigcup\limits_{d,d' \in \mathcal{D}} \{ w \in \mathcal{W} :y(d') > y(d),~u \in \mathcal{U}_{\{d,d'\}} \}
\end{align*}
with $\mathcal{U}_{\{d,d'\}} = \{u \in \mathcal{U} : d(u,\{d,d'\}) = d \}$.

\newpage
\bibliography{references.bib}

%% file: HSsumm.tex
 \begin{tabular}{ L{5cm}  C{3.5cm} C{3.5cm} }  
 \toprule  
   \multicolumn{3}{c}{Panel (a): Proportion in each care setting by experimental group} \\ \cline{1-3}  
 Care setting & Treatment & Control \\   \hline  
 Home care &  0.093 & 0.542  \\  
 Alternative preschool &  0.082 & 0.318  \\  
 Head Start &  0.825 & 0.140  \\ \hline  
 Sample size &  2290 & 1337  \\   
 \midrule  
   \multicolumn{3}{c}{Panel (b): ITT and IV Estimands} \\ \cline{1-3}  
 Estimand & Discretized & Raw \\   \hline  
 $\text{ITT}^D$ &  \multicolumn{2}{c}{0.685}  \\  
           &  \multicolumn{2}{c}{(0.018)}  \\ \cline{2-3}  
 $\text{ITT}^Y$ &  0.170 & 0.151  \\ 
           &  (0.035) & (0.030)  \\ \cline{2-3}  
 IV &  0.249 & 0.220  \\  
           &  (0.051) & (0.044)  \\  
 \bottomrule 
 \end{tabular}

%% file: HSmain.tex
 \begin{tabular}{ L{4.25cm}  R{1.5cm} R{1.5cm} R{1.5cm} R{1.5cm} R{1.5cm} R{1.5cm} R{1.5cm} R{1.5cm}}  
 \toprule  
 Assumption &  \multicolumn{1}{c}{(1)} & \multicolumn{1}{c}{(2)} & \multicolumn{1}{c}{(3)} & \multicolumn{1}{c}{(4)} & \multicolumn{1}{c}{(5)} & \multicolumn{1}{c}{(6)} & \multicolumn{1}{c}{(7)} & \multicolumn{1}{c}{(8)} \\ \hline  
 NC        &            &  \multicolumn{1}{c}{\checkmark} & \multicolumn{1}{c}{\checkmark} & \multicolumn{1}{c}{\checkmark} & \multicolumn{1}{c}{\checkmark} & \multicolumn{1}{c}{\checkmark} & \multicolumn{1}{c}{\checkmark} & \multicolumn{1}{c}{\checkmark}  \\  
 MTR($\lambda$)        &  & &  \multicolumn{1}{c}{1} & \multicolumn{1}{c}{0.9} & \multicolumn{1}{c}{0.8} &   &     &      \\  
 Roy($\lambda$)        &  & &    &     &     & \multicolumn{1}{c}{1} & \multicolumn{1}{c}{0.9} & \multicolumn{1}{c}{0.8}  \\ [1ex]  
 \midrule  
 Parameter   & & & & &  & & &  \\ \hline  
 \multirow{4}{*}{$\text{ATE}^D(C_{+p},C_{-p})$} & \scriptsize{0.795} & \scriptsize{0.795} & \scriptsize{0.795} & \scriptsize{0.795} & \scriptsize{0.795} & \scriptsize{0.795} & \scriptsize{0.795} & \scriptsize{0.795}  \\  
                                  & \multirow{2}{*}{0.825} & \multirow{2}{*}{0.825} & \multirow{2}{*}{0.825} & \multirow{2}{*}{0.825} & \multirow{2}{*}{0.825} & \multirow{2}{*}{0.825} & \multirow{2}{*}{0.825} & \multirow{2}{*}{0.825} \\  
                                  &   &   &   &   &   &   &   &   \\  
                                  & \scriptsize{0.855} & \scriptsize{0.855} & \scriptsize{0.855} & \scriptsize{0.855} & \scriptsize{0.855} & \scriptsize{0.855} & \scriptsize{0.855} & \scriptsize{0.855}   \\     
 \cline{2-9} 
  \multirow{4}{*}{$\text{ATE}^Y(C_{+p},C_{-p})$} & \scriptsize{-0.203} & \scriptsize{-0.203} & \scriptsize{0.060} & \scriptsize{0.048} & \scriptsize{0.006} & \scriptsize{0.060} & \scriptsize{0.048} & \scriptsize{0.006}  \\  
                                  & -0.113 & -0.113 & 0.170 & 0.118 & 0.076 & 0.170 & 0.118 & 0.076 \\  
                                  & 0.410 & 0.410 & 0.410 & 0.410 & 0.410 & 0.410 & 0.410 & 0.410 \\  
                                  & \scriptsize{0.500} & \scriptsize{0.500} & \scriptsize{0.510} & \scriptsize{0.500} & \scriptsize{0.500} & \scriptsize{0.510} & \scriptsize{0.500} & \scriptsize{0.500}   \\     
 \cline{2-9} 
  \multirow{4}{*}{$\text{ATOP}(C_{+p},C_{-p})$} & \scriptsize{-0.247} & \scriptsize{-0.247} & \scriptsize{0.076} & \scriptsize{0.053} & \scriptsize{0.002} & \scriptsize{0.076} & \scriptsize{0.053} & \scriptsize{0.002}  \\  
                                  & -0.137 & -0.137 & 0.206 & 0.143 & 0.092 & 0.206 & 0.143 & 0.092 \\  
                                  & 0.497 & 0.497 & 0.497 & 0.497 & 0.497 & 0.497 & 0.497 & 0.497 \\  
                                  & \scriptsize{0.617} & \scriptsize{0.617} & \scriptsize{0.617} & \scriptsize{0.617} & \scriptsize{0.617} & \scriptsize{0.617} & \scriptsize{0.617} & \scriptsize{0.617}   \\     
 \bottomrule 
 \end{tabular}

%% file: HShetero.tex
 \begin{tabular}{ L{4.25cm}  R{1.5cm} R{1.5cm} R{1.5cm} R{1.5cm} R{1.5cm} R{1.5cm} R{1.5cm} R{1.5cm}}  
 \toprule  
    & \multicolumn{8}{c}{Specification} \\ \cline{2-9}  
 Parameter   &  \multicolumn{1}{c}{(1)} & \multicolumn{1}{c}{(2)} & \multicolumn{1}{c}{(3)} & \multicolumn{1}{c}{(4)} & \multicolumn{1}{c}{(5)} & \multicolumn{1}{c}{(6)} & \multicolumn{1}{c}{(7)} & \multicolumn{1}{c}{(8)}  \\ \hline  
 \multirow{4}{*}{$\text{ATE}^D_{+a}(C_{+p},C_{-p})$} & \scriptsize{0.515} & \scriptsize{0.515} & \scriptsize{0.505} & \scriptsize{0.515} & \scriptsize{0.515} & \scriptsize{0.505} & \scriptsize{0.515} & \scriptsize{0.515}  \\  
                                  & 0.575 & 0.575 & 0.575 & 0.575 & 0.575 & 0.575 & 0.575 & 0.575  \\  
                                  & 0.910 & 0.822 & 0.822 & 0.822 & 0.822 & 0.822 & 0.822 & 0.822  \\  
                                  & \scriptsize{0.930} & \scriptsize{0.862} & \scriptsize{0.872} & \scriptsize{0.862} & \scriptsize{0.862} & \scriptsize{0.872} & \scriptsize{0.862} & \scriptsize{0.862}  \\ \cline{2-9}    
 \multirow{4}{*}{$\text{ATOP}_{+a}(C_{+p},C_{-p})$} & \scriptsize{-1.572} & \scriptsize{-1.514} & \scriptsize{0.000} & \scriptsize{-0.374} & \scriptsize{-0.658} & \scriptsize{0.000} & \scriptsize{-0.374} & \scriptsize{-0.658}  \\  
                                  & -1.392 & -1.324 & 0.000 & -0.351 & -0.578 & 0.000 & -0.351 & -0.578  \\  
                                  & 1.554 & 1.415 & 1.028 & 1.292 & 1.397 & 1.028 & 1.292 & 1.397  \\  
                                  & \scriptsize{1.764} & \scriptsize{1.625} & \scriptsize{1.308} & \scriptsize{1.562} & \scriptsize{1.637} & \scriptsize{1.308} & \scriptsize{1.562} & \scriptsize{1.637}  \\ \cline{2-9}    
 \hline 
 \multirow{4}{*}{$\text{ATE}^D_{-a}(C_{+p},C_{-p})$} & \scriptsize{0.000} & \scriptsize{0.788} & \scriptsize{0.788} & \scriptsize{0.788} & \scriptsize{0.788} & \scriptsize{0.788} & \scriptsize{0.788} & \scriptsize{0.788}  \\  
                                  & \multirow{2}{*}{-} & 0.828 & 0.828 & 0.828 & 0.828 & 0.828 & 0.828 & 0.828  \\  
                                  &   & 1.000 & 1.000 & 1.000 & 1.000 & 1.000 & 1.000 & 1.000  \\  
                                  & \scriptsize{1.000} & \scriptsize{1.000} & \scriptsize{1.000} & \scriptsize{1.000} & \scriptsize{1.000} & \scriptsize{1.000} & \scriptsize{1.000} & \scriptsize{1.000}  \\    
 \cline{2-9} 
  \multirow{4}{*}{$\text{ATOP}_{-a}(C_{+p},C_{-p})$} & \scriptsize{-3.739} & \scriptsize{-0.771} & \scriptsize{0.022} & \scriptsize{-0.239} & \scriptsize{-0.378} & \scriptsize{0.022} & \scriptsize{-0.239} & \scriptsize{-0.378}  \\  
                                  & \multirow{2}{*}{-} & -0.641 & 0.072 & -0.169 & -0.298 & 0.072 & -0.169 & -0.298  \\  
                                  &   & 1.061 & 0.717 & 0.856 & 0.930 & 0.717 & 0.856 & 0.930  \\  
                                  & \scriptsize{3.739} & \scriptsize{1.221} & \scriptsize{0.907} & \scriptsize{1.036} & \scriptsize{1.100} & \scriptsize{0.907} & \scriptsize{1.036} & \scriptsize{1.100}  \\    
 \bottomrule 
 \end{tabular}

%% file: HScostbenefit.tex
 \begin{tabular}{ L{2cm}  R{1.1cm} R{1.1cm} R{.025cm} R{1.1cm} R{1.1cm} R{.025cm} R{1.1cm} R{1.1cm} R{.025cm} R{1.1cm} R{1.1cm} R{.025cm} R{1.1cm} R{1.1cm} R{.025cm} R{1.1cm} R{1.1cm}}  
 \toprule  
 & \multicolumn{17}{c}{Specification} \\ \cline{2-18}  
 & \multicolumn{2}{c}{(4)} & & \multicolumn{2}{c}{(5)} & & \multicolumn{2}{c}{(6)} & & \multicolumn{2}{c}{(7)} & & \multicolumn{2}{c}{(8)} & & \multicolumn{2}{c}{(9)} \\   
 & \multicolumn{2}{c}{$\kappa$} & & \multicolumn{2}{c}{$\kappa$} & & \multicolumn{2}{c}{$\kappa$} & & \multicolumn{2}{c}{$\kappa$} & & \multicolumn{2}{c}{$\kappa$} & & \multicolumn{2}{c}{$\kappa$} \\  \cline{2-3} \cline{5-6} \cline{8-9} \cline{11-12} \cline{14-15} \cline{17-18}  
 Parameter & \multicolumn{1}{c}{0.1} & \multicolumn{1}{c}{0.3} & & \multicolumn{1}{c}{0.1} & \multicolumn{1}{c}{0.3} & & \multicolumn{1}{c}{0.1} & \multicolumn{1}{c}{0.3} & & \multicolumn{1}{c}{0.1} & \multicolumn{1}{c}{0.3} & & \multicolumn{1}{c}{0.1} & \multicolumn{1}{c}{0.3} & & \multicolumn{1}{c}{0.1} & \multicolumn{1}{c}{0.3} \\ \hline  
 \multirow{4}{*}{$\text{AC}$} & \scriptsize{3,946} & \scriptsize{3,946} & & \scriptsize{3,946} & \scriptsize{3,946} & & \scriptsize{3,946} & \scriptsize{3,946} & & \scriptsize{3,946} & \scriptsize{3,946} & & \scriptsize{3,946} & \scriptsize{3,946} & & \scriptsize{3,946} &  \scriptsize{3,946}  \\  
                                  & 4,346 & 4,346 & & 4,346 & 4,346 & & 4,346 & 4,346 & & 4,346 & 4,346 & & 4,346 & 4,346 & & 4,346 & 4,346  \\  
                                  & 5,185 & 5,185 & & 5,185 & 5,185 & & 5,185 & 5,185 & & 5,185 & 5,185 & & 5,185 & 5,185 & & 5,185 & 5,185  \\  
                                  & \scriptsize{5,585} & \scriptsize{5,585} & & \scriptsize{5,585} & \scriptsize{5,585} & & \scriptsize{5,585} & \scriptsize{5,585} & & \scriptsize{5,585} & \scriptsize{5,585} & & \scriptsize{5,585} & \scriptsize{5,585} & & \scriptsize{5,585} & \scriptsize{5,585}    \\   
 \cline{2-18}   
  \multirow{4}{*}{$\text{AB}$} & \scriptsize{2,250} & \scriptsize{6,750} & & \scriptsize{1,654} & \scriptsize{5,163} & & \scriptsize{208} & \scriptsize{625} & & \scriptsize{2,250} & \scriptsize{6,750} & & \scriptsize{1,654} & \scriptsize{5,163} & & \scriptsize{208} &  \scriptsize{625}  \\  
                                  & 5,850 & 17,550 & & 4,054 & 12,163 & & 2,608 & 7,825 & & 5,850 & 17,550 & & 4,054 & 12,163 & & 2,608 & 7,825  \\  
                                  & 14,073 & 42,219 & & 14,073 & 42,219 & & 14,073 & 42,219 & & 14,073 & 42,219 & & 14,073 & 42,219 & & 14,073 & 42,219  \\  
                                  & \scriptsize{17,473} & \scriptsize{52,419} & & \scriptsize{17,273} & \scriptsize{51,419} & & \scriptsize{17,273} & \scriptsize{51,419} & & \scriptsize{17,473} & \scriptsize{52,419} & & \scriptsize{17,273} & \scriptsize{51,419} & & \scriptsize{17,273} & \scriptsize{51,419}    \\   
 \cline{2-18}   
  \multirow{4}{*}{$\text{AS}$} & \scriptsize{-2,935} & \scriptsize{1,565} & & \scriptsize{-3,531} & \scriptsize{-22} & & \scriptsize{-5,177} & \scriptsize{-4,560} & & \scriptsize{-2,935} & \scriptsize{1,565} & & \scriptsize{-3,531} & \scriptsize{-1,022} & & \scriptsize{-5,177} &  \scriptsize{-5,360}  \\  
                                  & 665 & 12,365 & & -1,131 & 6,978 & & -2,577 & 2,640 & & 665 & 12,365 & & -1,131 & 6,978 & & -2,577 & 2,640  \\  
                                  & 9,727 & 37,873 & & 9,727 & 37,873 & & 9,727 & 37,873 & & 9,727 & 37,873 & & 9,727 & 37,873 & & 9,727 & 37,873  \\  
                                  & \scriptsize{13,127} & \scriptsize{48,073} & & \scriptsize{12,927} & \scriptsize{47,273} & & \scriptsize{12,927} & \scriptsize{47,273} & & \scriptsize{13,127} & \scriptsize{48,073} & & \scriptsize{12,927} & \scriptsize{47,273} & & \scriptsize{12,927} & \scriptsize{47,273}    \\   
 \bottomrule 
 \end{tabular}